\documentclass[12pt,dvips]{article}

\usepackage{amsmath,amssymb,exscale}
\usepackage{array,multicol}
\usepackage{afterpage,float,flafter}
\usepackage{epsfig,rotating,pifont}
\usepackage{cite}
\@addtoreset{equation}{section} \makeatother

\def\beq{\begin{eqnarray}}
\def\eeq{\end{eqnarray}}

\def\lsim{\mathrel{\rlap{\lower3pt\hbox{\hskip0pt$\sim$}}
  \raise1pt\hbox{$<$}}}         
\def\gsim{\mathrel{\rlap{\lower4pt\hbox{\hskip1pt$\sim$}}
  \raise1pt\hbox{$>$}}}         

\title{
\vspace{0cm}
\huge{Gauge Boson Mass Generation in AdS$_4$}
\vspace*{0.7cm}
\author{
\Large {\text{Riccardo Rattazzi and Michele Redi}}\\ \\
\emph{Institut de Th\'eorie des Ph\'enom\`enes Physiques}\\
\emph{EPFL, CH-1015, Lausanne, Switzerland}
}
}

\date{}
\begin{document}
\maketitle \thispagestyle{empty} \vspace*{-.2cm}

\begin{abstract}
We investigate the role of boundary conditions in gauge theories in AdS$_4$.
The presence of the boundary can break the gauge symmetry consistently with AdS$_4$ isometries.
We show that, as a consequence, the gauge bosons associated to the broken symmetries
become massive at one loop. In particular chiral gauge theories such us the Standard Model
are necessarily massive in AdS$_4$. We briefly discuss similarities with the Schwinger
model and implications for CFTs in three dimensions.

\end{abstract}

\newpage
\renewcommand{\thepage}{\arabic{page}}
\setcounter{page}{1}

\section{Introduction}

Symmetry principles play an essential role in constraining the spectrum of quantum systems.
In particular, in quantum field theory,  massless particles are often understood as a
consequence of symmetry. For spin 0 and spin 1/2 particles, supersymmetry and chiral symmetry
are the relevant principles. For particles of spin 1 and higher, gauge symmetry is the relevant
invariance. However gauge symmetry is not an ordinary symmetry but  a redundancy in the
parametrization of the dynamical variables. In the simplest situations this redundancy
ensures the absence of the additional physical polarizations that are necessary to
endow with mass particles with spin $\geq 1$. This is the case of QED in 4D.
In more general situations the gauge symmetry is not enough to forbid degrees of freedom
acting as the extra polarizations, and then mass generation follows. In this case the gauge
theory is said to be in the Higgs phase, and the additional polarizations are associated to
Nambu-Goldstone (NG) bosons non-linearly realizing the gauge symmetry. In weakly coupled gauge
theories the NG-bosons are elementary states. An example of that is given by the Standard Model
with an elementary Higgs field. On the other hand if the interaction is sufficiently strong
the role of NG-bosons can be played by composite states. This situation is realized for instance
in technicolor models in 4D. Heuristically, a strong interaction among elementary constituents
is needed in order to produce a NG pole out of a perturbative continuum spectrum in the
current-current correlator.

The purpose of this paper is to illustrate how a Higgs mechanism involving a NG-boson composed by
two elementary particles can arise at the perturbative level in gauge theories on AdS$_4$.
The geometry  of AdS$_4$ is crucial for this phenomenon to happen. On one side,  in AdS$_4$ energy levels
are discrete much like in finite volume (although AdS$_4$ has infinite volume),
and therefore multi-particles states have a discrete mass spectrum. Moreover, since null
geodesics reach the boundary of AdS$_4$ in finite time, bulk physics is crucially affected
by boundary conditions. Indicating by $D(E,s)$ the one particle representations of the AdS
algebra \cite{nicolai} whose ground state has energy $E$ and spin $s$, our basic point
is the following. Measuring energies in units of the inverse AdS radius, a massless
fermion $\psi$ corresponds to  $D(\frac{3}{2},\frac{1}{2})$.  The two particle Hilbert
space $\psi \otimes \psi$ then obviously contains the scalar representation $D(3,0)$.
This corresponds to a derivatively coupled 4D scalar, a candidate NG-boson. Whether and
how this scalar shifts under a gauge symmetry and thus causes the associated vector field
to acquire a mass depends on the boundary conditions. Our basic remark is that
charge breaking boundary conditions are compatible with AdS isometries. When charge
breaking boundary conditions are imposed, the two fermion composite state is eaten by
the bulk vector giving rise to a massive spin 1 multiplet
\begin{equation}
D\left (2,1\right )\oplus
D\left (3,0\right ) \longrightarrow D\left (\frac{3+\sqrt{1+4m^2}}{2},1\right )
\end{equation}
As the mass is due to the mixing between 1- and 2-particle states, it will arise by
considering the vector self-energy at 1-loop: $m^2\sim \frac{\alpha}{4\pi}$.
A similar phenomenon can also arise for conformally coupled bulk scalars $\phi$, that,
depending on boundary conditions \cite{bf,klebanovwitten}, can be quantized as either $D(1,0)$ or $D(2,0)$.
Again, in general we shall have $D(3,0)\subset \phi\otimes \phi$. In the case of scalars there
is always the option to choose charge preserving boundary conditions. On the contrary,
for chiral gauge theories the boundary necessarily breaks the gauge group to a subgroup
of the maximal vector subgroup. So, for instance, in the Standard Model on AdS$_4$ even
in the absence of an elementary Higgs field, the electro-weak vector bosons have a small mass.

This paper is organized as follows. In Sect. 2 we introduce general boundary conditions
for massless fermions and compute the 1-loop contribution to the vector boson mass.
Furthermore we discuss the result for QED and for chiral gauge theories and
illustrate how things change for massive fermions. In Sect. 3 we include scalars and
extend our result to supersymmetric gauge theories. We check that the resulting vector
and gaugino masses are consistent with the Super-AdS algebra. In particular
the so called `anomaly mediated' gaugino mass is an essential contribution.
Finally, in Sect. 4 we discuss our results comparing to the mass generation
in the Schwinger model in 1+1 dimensions and providing a holographic interpretation
according to the AdS/CFT correspondence.

\section{Mass Generation}

Let us consider a gauge theory with group $G$ in AdS$_4$ space coupled to $n$ massless Weyl fermions
in a representation of $G$,  generally reducible and anomaly free. The bulk action reads
\begin{eqnarray}
S&=&\int d^3x dz\,e \left[-\frac 1 2 Tr F^2 -\frac i 2 (\bar{\psi}^i \bar{\sigma}^M D_M \psi_i +h.c.)\right]\nonumber \\
&& D_M =\partial_M+\omega_M+i g  A_M^a T_a
\end{eqnarray}
where $\omega_M$ is the spin connection.
Working in the Poncair\'e patch we take $e_M^A=L/z\,\delta_M^A$ and we follow all the conventions of
\cite{wb}. In the massless limit the action is Weyl invariant so that it can be rescaled (classically)
to half of flat space. It turns out that, for the computation we are interested in, we can easily bypass the
complication associated to the breaking of Weyl rescaling by the UV regulator. A similiar approach was
taken in ref.\cite{gaugino}. We will therefore perform all the computations using flat space variables.

In AdS the presence of a boundary at $z=0$ requires that boundary terms are added
to the action in order to make the variational problem well defined. These terms are in general described by a symmetric matrix $B$
\begin{equation}
\frac 1 4 \int d^3 x\, B_{ij}\,\psi^i \psi^j+h.c.
\end{equation}
implying the following boundary conditions
\begin{equation}
\psi_i|_{z=0}=- i B_{ij}^* \sigma^3 \bar{\psi}^j|_{z=0}.
\label{bcfermion}
\end{equation}
The existence of non trivial ($\psi\not =0$) solutions to the above equation requires that $B$ is a symmetric unitary matrix.
In the massless limit the free bulk matter lagrangian has a chiral symmetry $U(n)$.
However, since the boundary matrix $B$ transforms under a chiral rotation as $B \to U^T  B U$,
it follows that the boundary breaks the symmetry to $O(n)$. This will necessarily break part of the
gauge symmetry unless $G\subseteq O(n)$. The unbroken generators of $G$ satisfy
\begin{equation}
T_a^* B + B T_a=0
\label{unbrokengenerators}
\end{equation}
i.e. they provide a real representation of the algebra.
Note that such a general form of mass matrix would not be allowed in the bulk:
`explicit' breaking of the gauge symmetry in the bulk is equivalent to adding
the corresponding elementary NG-bosons, in contradiction with the goal stated
in the Introduction. In AdS space, however, the fields at the boundary are not dynamical and charge
breaking conditions can be imposed. By eq.~(\ref{bcfermion}) there is no energy-momentum
flow at the boundary, thus ensuring the compatibility with the isometries of AdS$_4$.
This property distinguishes our set up from previous literature on mass generation
on AdS$_4$, where transparent boundary conditions were imposed, corresponding to the presence of extra states (associated to a defect CFT) \cite{Porrati:2001db,burrington}. In our set up only the gauge charge flows through the boundary. Indeed it must be stressed
that, for chiral gauge theories, charge breaking at the boundary is mandatory, as the representation is complex and only a subgroup of
$G$ can be preserved. The necessity to break chirality in AdS$_4$ as a consequence of the relevance of the 3D boundary
was first noticed in ref.\cite{allen}, but the implications for chiral gauge theories where not investigated in that paper.

The above boundary conditions  determine the fermion propagators to be
\begin{eqnarray}
\langle
\psi_{i\alpha}(X_1)\bar{\psi}_{j\dot\beta}(X_2)\rangle&
= &
\frac i {2 \pi^2} \frac{(X_1-X_2)_M
\sigma^M_{\alpha\dot{\beta}}}
{[(X_1-X_2)^2+i\epsilon ]^2}\, \delta_{ij},\label{direct} \\
\langle
\psi_{i\alpha}(X_1){\psi}_{j}^{\beta}(X_2)\rangle& =
&-\frac 1{2 \pi^2} \frac{(X_1-\tilde{X}_2)_M (\sigma^M\bar{\sigma}^3)_\alpha^\beta}
{[(X_1-\tilde{X}_2)^2+i\epsilon ]^2}\, B_{ij}^*.
\label{reflected}
\end{eqnarray}
where $\tilde{X}=(x,-z)$. Eqs. (\ref{direct}) and (\ref{reflected}) are naturally associated to, respectively, direct propagation  and propagation with one reflection at the boundary. The second equation implies the presence of a condensate in the bulk,
\begin{equation}
\langle\psi_i \psi_j(X)\rangle= \frac 1 {8\pi^2} \frac {B_{ij}^*}{z^3}
\label{condensate}
\end{equation}
which spontaneously breaks the chiral $U(n)$ global symmetry of the bulk action to $O(n)$.

We are now ready to compute the gauge boson mass at 1-loop. The self-energy due to the matter action
decomposes into two contributions, from direct and reflected propagators\footnote{We will not compute the contribution from gauge loops as, first of all, it is independent of the matter contribution and  absent is abelian theories. Secondly,  we could not identify any two particle state playing the role of the NG-boson in the two vector channel.}. Let us consider first the second contribution which is not ambiguous and does not require
regularization. Defining the 1PI effective action as $\Gamma_{1PI}\in \int 1/2 A_M^a(X_1) \Pi^{MN}_{ab}(X_1,X_2) A_{Nb}(X_2)$
we have,
\begin{eqnarray}
\Pi^{MN}_{R\,ab}(X_1,X_2)&=&-i\, g^2 \kappa_{a b} \left(\frac i {2\pi^2}\right)^2 Tr\left[\sigma^M\bar{\sigma}^3\sigma^Q\bar{\sigma}^N\sigma^P\bar{\sigma}^3\right]\frac {(X_1-\tilde{X}_2)_P(X_1-\tilde{X}_2)_Q}{(X_1-\tilde{X}_2)^8}\nonumber\\ &=&-i\, \frac {g^2 \kappa_{ab}} {2\pi^4} \left[\frac {\tilde{\eta}^{MN}}{(X_1-\tilde{X}_2)^6}-\frac {2 (X_1-\tilde{X}_2)^M(X_1-\tilde{X}_2)^P\tilde{\eta}_P^N}{(X_1-\tilde{X}_2)^8}\right]
\end{eqnarray}
where $\tilde{\eta}_{MN}=Diag(-1,1,1,-1)$ and $\kappa_{ab}=Tr[BT_a B^*T_b^*]$.
As expected $\Pi^{MN}_R$ is transverse, $\partial_{X_1^M} \Pi^{MN}_{R}(X_1,X_2)=\partial_{X_2^N} \Pi^{MN}_R(X_1,X_2)=0$, guaranteeing that the effective action is gauge invariant in the bulk.
Indeed the self-energy can be written as
\begin{equation}
\Pi^{MN}_{R\,ab}(X_1,X_2)=i \frac 2 3 \frac {g^2 \kappa_{ab}} {(4\pi^2)^2} \tilde{\eta}^M_P\left(\frac {\partial}{\partial X_{2N}}\frac {\partial}{\partial X_{2P}}-\eta^{PN}\square_2\right)  \frac 1
{(X_1-\tilde{X}_2)^4}
\end{equation}

To extract the photon mass we proceed as in \cite{gaugino} and compute the 1-loop corrected
`equations of motion' associated to the 1PI effective action. We suppress the non abelian indices as they  factor out.
By integrating twice by parts the 1-loop contribution to the equations of motion, we obtain
\begin{eqnarray}
EQM_R&=&\int d^4X_2\, \Pi_{R}^{MN}(X_1,X_2) A_N(X_2)\nonumber \\
&=&-i\frac 2 3 \frac {g^2\kappa}{(4\pi^2)^2}\tilde{\eta}^{MP} \int d^4X_2 \frac 1{(X_1-\tilde{X}_2)^4} \frac {\partial}{\partial X_{2Q}} F_{QP}(X_2)\nonumber \\
&-& i\frac 2 3 \frac {g^2\kappa}{(4\pi^2)^2}\tilde{\eta}^{MP}  \int d^3x_2 \left[\eta_{P3} \frac {\partial}{\partial X_{2Q}} \frac 1{(X_1-\tilde{X}_2)^4} A_Q(X_2)-\frac {\partial}{\partial z_2} \frac 1{(X_1-\tilde{X}_2)^4} A_P(X_2)\right]\Big|_{z_2=0}
\nonumber \\
&-& i\frac 2 3 \frac {g^2\kappa}{(4\pi^2)^2}\tilde{\eta}^{MP}  \int d^3x_2 \left[\frac 1{(X_1-\tilde{X}_2)^4} F_{3P}(X_2)\right]\Big|_{z_2=0}
\label{eqmr}
\end{eqnarray}
At leading order the contribution to the mass can be derived by evaluating the equations of motion
on massless solutions. We find it convenient to use solutions that satisfy the gauge condition
\begin{equation}
\partial^M \left[\frac {A_M}{z^2}\right]=0
\label{gaugecondition}
\end{equation}
which corresponds, in our choice of coordinates,  to the general covariant Lorentz gauge condition $D^MA_M=0$.
For the same reasons explained in \cite{gaugino}, we also need to impose
Hartle-Hawking boundary conditions at the horizon $z=\infty$ of the Poincar\'e patch.
A physical (not pure gauge) set of solutions of the massless equations in Lorentz gauge is
\begin{equation}
A_3=0, \quad\quad A_\mu= e^{i(p_\nu x^\nu+|p|z)}\epsilon_\mu~~~~~~~~~~~~~~\epsilon_\mu p^\mu=0,~~~~~~~~~~~~~~\mu=0,1,2
\label{simplesolution}
\end{equation}
so we will compute the action of the self-energy on these functions and show it acts like a local mass term. Notice that in the massless case the Lorentz gauge leaves one residual gauge degree of freedom. This additional polarization is pure gauge in the massless case
but becomes the physical 3rd polarization in the massive case. We have checked that the self energy acts like the same local mass term
also on this additional polarization, for which computations are slightly more involved.

The bulk contribution in the second line of (\ref{eqmr}) is zero by the tree level equations of motion.
The boundary terms give rise to a mass term. To see this, the first term in the third line of
eq. (\ref{eqmr}) vanishes due to $\epsilon_\mu p^\mu=0$ while the second gives
\begin{equation}
i\frac 2 3\frac {g^2\kappa} {(4\pi^2)^2} \frac {\partial}{\partial z_1} \int  d^3 x_2
\frac{1}{[(x_1-x_2)^2+z_1^2+i\epsilon ]^2}\, e^{i\,  p_\nu x^\nu_2}\epsilon_\mu=-\frac 2 3\frac {g^2\kappa} {(4\pi)^2}\left[\frac 1 {z_1^2}-i \frac {|p|}{z_1}\right]\,  e^{i(p_\nu x^\nu_1+|p|z_1)}\epsilon_\mu.
\end{equation}
From the fourth line we obtain
\begin{equation}
-i\frac 2 3\frac {g^2\kappa} {(4\pi^2)^2}  \int  d^3 x_2
\frac{i |p|}{[(x_1-x_2)^2+z_1^2+i\epsilon ]^2}\,  e^{i\,p_\mu x^\mu_2}\epsilon_\mu=-\frac 2 3\frac {g^2\kappa} {(4\pi)^2}\left[i \frac {|p|}{z_1}\right]\, e^{i(p_\nu x^\nu_1 +|p|z_1)}\epsilon_\mu.
\end{equation}
Combining the two we see that $\Pi^{MN}_R$ acts as a mass term,
\begin{equation}
\delta m^2_{ab}\vert_{reflected}=\frac 2 {3L^2}\frac {g^2\kappa_{ab}} {(4\pi)^2}
\end{equation}

The contribution to the self-energy from the direct propagators is  more subtle, as it requires UV regulation. However
its computation can be bypassed. Indeed   the direct contribution to the vector boson self-energy  is proportional to ${\rm Tr}[T_a T_b]$, and
independent of the boundary matrix $B$. So we simply have
  \begin{equation}
 \delta m^2_{ab}\vert_{direct}=c\frac 1 {L^2}\frac {1}{(4\pi)^2} {\rm Tr}[T_a T_b]\,.
 \end{equation}
with $c$ a numerical coefficient. The request that the sum of reflected and direct contributions
to the vector mass vanish in the charge preserving case, $BT_a=-T^*_aB$,  fixes  $c=2/3$.
In the end the total contribution to the vector mass is
\begin{equation}
m^2_{ab}=\frac 2 {3L^2} \frac {g^2} {(4\pi)^2}\, {\rm {Tr}}[BT_aB^*T_b^*+T_a T_b]
\label{mass}
\end{equation}
As a matter of fact we have also performed an explicit computation of the direct contribution to the self-energy confirming
the above result. The matrix in eq.~~(\ref{mass}) is easily proven to be positive semi-definite.
Indeed $T\to B^*T^*B\equiv T_B$ is an orthogonal transformation with respect to the natural metric ${\rm {Tr}} T_1T_2\equiv \langle T_1|T_2\rangle$  on the space of hermitean matrices.  Defining $T=\alpha_a T_a$ we have then
\begin{equation}
\alpha_a\alpha_b m^2_{ab}\propto \langle T|T_B\rangle+ \langle T|T\rangle\geq 0
\end{equation}
with $\alpha_a\alpha_b m^2_{ab}=0$ occurring if and only if $T=- B^*T^*B\equiv T_B$.
Therefore the gauge bosons associated to the broken generators acquire positive mass$^2$.
This is the main result of our paper.

\subsection{QED}

To be concrete, let us consider QED coupled to two massless Weyl fermions of opposite charges. In this
case the general boundary matrix depends on three real parameters and can be conveniently parameterized as,
\begin{eqnarray}
B = \left(
\begin{array}{cc}
i \lambda\, e^{2 i \phi_1} & \sqrt{1-\lambda^2}\, e^{i (\phi_1+\phi_2)}\\
\sqrt{1-\lambda^2}\, e^{i (\phi_1+\phi_2)} & i \lambda \,e^{2 i  \phi_2}  \\
\end{array}
\right)
\label{Bgeneral}
\end{eqnarray}
with $0\leq \lambda\leq 1$.
From eq. (\ref{mass}) the mass of the photon is $m_{\gamma}^2=g^2\,\lambda^2/(6\pi^2 L^2)$
and does not depend on $\phi_{1,2}$. Indeed the bulk Lagrangian of massless QED is classically invariant under
a $U(1)\times U(1)$ symmetry corresponding to electric charge and chiral symmetry. Due to this bulk
symmetry, the two phases can be eliminated by the a field redefinition that does affect the bulk lagrangian.
Note however that by adding bulk operators that break the chiral symmetry (for instance 4 fermion interactions)
the combination $\phi_1+\phi_2$ becomes observable, while $\phi_1-\phi_2$ can always be set to zero by charge rotations.
In the end, the physically relevant parameters are in general 2, with the vector boson mass taking values in a fixed range $0\leq m_\gamma^2L^2\leq g^2/6\pi^2$. Notice that in the general case of many fermions with charge matrix $Q$, the maximal value the vector mass can attain is proportional to  ${\rm Tr}Q^2$, that is the same combination of charges that controls the 1-loop $\beta$-function.

\subsection{Chiral Theories}

As we mentioned earlier, when the matter fields are in a complex representation of
$G$, i.e. the gauge theory is chiral, no boundary conditions that preserve the full gauge symmetry
can be chosen and therefore some of the gauge bosons necessarily become massive.
In this case the maximal symmetry that can be preserved is the maximal
vectorial subgroup of $G$.

As an example let us consider an $SU(5)$ gauge theory with fermions in the $5+\bar{10}$ representation.
The maximal vector subgroup is $SU(4)$ under which the fermions decompose as $1\oplus 4 \oplus \bar{4} \oplus 6$.
The broken generators transform in the $4\oplus \bar{4}\oplus 1$ representation of $SU(4)$ and the associated gauge bosons
will acquire mass. For the Standard Model the maximal vectorial subgroup is instead $SU(3)\otimes U(1)$. Let us consider a quark doublet.
In order the preserve $SU(3)\otimes U(1)$ the boundary conditions give rise to the following condensates,
\begin{eqnarray}
<d d_c>&=& \frac 3 {8\pi^2 z^3}\nonumber \\
<u u_c>&=& \frac 3 {8\pi^2 z^3}
\end{eqnarray}
The pattern of chiral symmetry breaking is identical to the one in QCD and the NG-bosons
associated to this breaking become the longitudinal components of the $W,Z$ bosons.
In fact, as in the Standard Model, due to the unbroken $SU(2)$ ``custodial'' symmetry rotating up
and down quarks, we have $m_{\scriptscriptstyle W}^2/m_{\scriptscriptstyle Z}^2=\cos^2 \theta_{\scriptscriptstyle W}$ with $\theta_{\scriptscriptstyle W}$ the Weinberg angle. The same conclusion
holds in the lepton sector if there exists a right-handed neutrino. Otherwise there will be necessarily a $\nu^2$ condensate
which breaks custodial symmetry (and lepton number) and modifies the previous ratio. Notice, finally, that in the $SU(5)$ and $SU(3)\times SU(2)\times U(1)$ examples the vector boson mass at the point of maximal symmetry is fixed.

\subsection{Bulk Mass}

For vector theories we can add a bulk mass for the fermions.
In this case the boundary conditions consistent with AdS$_4$ invariance are more restricted: the bulk mass
gives a discrete set of possibilities for the scaling of the solution at the boundary. This follows from the fact that the boundary matrix
and the bulk mass matrix must be simultaneously diagonalizable.
As discussed in \cite{gaugino}, for a Weyl spinor of mass $mL\ge 1/2$ (we assume without loss of
generality $m$ to be real and positive) the bulk action already implies the boundary condition uniquely. The resulting
one particle Hilbert space corresponds to the $D(\frac{3}{2}+mL,\frac{1}{2})$ representation.
In the case of QED this unique boundary condition is, not surprisingly charge preserving, implying a massless photon.
As a quick check of that, notice that the tensor product $D(\frac{3}{2}+mL,\frac{1}{2})\otimes D(\frac{3}{2}+mL,\frac{1}{2})$ does
not contain the NG representation $D(3,0)$. In the region $mL<1/2$, analogously to the scalar double quantization \cite{bf,klebanovwitten},
two inequivalent boundary conditions are allowed for each Weyl fermion. In the presence of multiple Weyl fermions with the same mass, the above discrete set of boundary conditions can be folded by a rotation among the fields of equal mass.
If one performs this exercise for QED coupled to a massive Dirac fermion, one finds 3
inequivalent possibilities for the matrix $B$ describing the boundary condition
\begin{equation}
B_\pm=
\pm \left(
\begin{array}{cc}
  0  &1  \\
 1   &   0
\end{array}
\right)
\quad\quad\qquad
B_0=
\left(
\begin{array}{cc}
  e^{ i\phi}  &0   \\
  0   &   e^{- i\phi}
\end{array}
\right)\, .
\end{equation}
The choices $B_+$ and $B_-$ preserve charge and correspond to a Dirac fermion in respectively  $D(\frac{3}{2}+mL,\frac{1}{2})$ and $D(\frac{3}{2}-mL,\frac{1}{2})$. The choice $B_0$ breaks charge maximally. The one particle Hilbert space corresponds to a direct sum of Majorana spinors $D(\frac{3}{2}+mL,\frac{1}{2})\oplus D(\frac{3}{2}-mL,\frac{1}{2})$. In this case, as expected, a Goldstone multiplet appears in the two particle Hilbert space. The angle $\phi$ has no physical consequences as it can be eliminated by a gauge rotation.
Not surprisingly, since chiral symmetry is broken by the bulk mass, the freedom in $\phi_1+\phi_2$ has disappeared with respect to eq.~(\ref{Bgeneral}). But, less trivially, the parameter $\lambda$ controlling charge breaking is now `quantized' to be either 0 or 1.
Contrary to the massless case, the action is not Weyl invariant so that flat
space formulae cannot be used to compute the vector mass and a genuine AdS$_4$ computation is required.
We of course expect the gauge boson to be massive also in this case
with a mass that goes to zero as $mL\to 1/2$.

\section{Supersymmetry}

The previous results can be easily extended to the supersymmetric version of the theory.
Supersymmetric QED in AdS$_4$ with charge preserving boundary
conditions was studied in \cite{gaugino} so we will consider this case.
In that paper it was shown that an ultraviolet counter-term, the anomaly mediated gaugino
mass \cite{anomalymediation}, was required to cancel an infrared contribution associated
to the $R-$symmetry breaking from the boundary in order to leave the gaugino massless
as demanded by supersymmetry.

In general, supersymmetry also allows for charge breaking boundary conditions
with the scalars in a chiral multiplet aligned with the fermions
\begin{equation}
\phi_i|_{z=0}= B_{ij}^* \phi^{j*}|_{z=0}.
\end{equation}
For zero mass term in the superpotential the scalars are also conformally coupled so
the full action can again be rescaled to half of flat space.
Up to a numerical factor, the contribution to the photon self energy from the scalar loop is identical
to the one of the fermions. In fact this loop is proportional to the $\beta-$function
of the theory. Since a complex scalar
contributes one half of a Weyl spinor in the $\beta-$function we find that the photon mass is
$3/2$ of eq. (\ref{mass}). In terms of AdS$_4$ representations (see \cite{nicolai}),
since for vector fields $m_1^2L^2=E(E-3)+2$ this corresponds to
\begin{equation}
D\left(2+\frac {g^2(\kappa+2)}{(4\pi)^2},1\right)\qquad\quad \kappa\equiv {\rm Tr} BQB^*Q^*
\end{equation}

By imposing charge breaking boundary conditions the infrared contribution to gaugino mass will not
cancel exactly the anomaly mediated one, which is independent of the boundary conditions.
Repeating the same steps as in \cite{gaugino} one finds
\begin{equation}
m_\lambda=\frac {g^2(k+2)}{(4\pi)^2L}
\end{equation}
This corresponds to the AdS representation ($m_{1/2}L=E-3/2$),
\begin{equation}
D\left(\frac 3 2+\frac {g^2(k+2)}{4\pi^2},\frac 1 2\right)
\end{equation}
Recalling that a massive vector multiplet decomposes into the following representations
\begin{equation}
D\left( E_0 ,\frac 1 2\right)\oplus D\left(E_0+
\frac 1 2 , 0\right)\oplus D\left(E_0+\frac 1 2 , 1\right)\oplus D\left (E_0+1,\frac 1 2 \right)\, ,
\end{equation}
we conclude that the photon and gaugino acquire masses as demanded by supersymmetry with $E_0=\frac 3 2+\frac {g^2(k+2)}{4\pi^2}$.
Notice that  $D(E_0+\frac 1 2 , 0)$ and $D(E_0+1,\frac 1 2 )$ correspond to purely two-particle states.
In order to check the satisfaction of the algebra for these states we would have to study the
K\"allen-Lehmann decomposition of the current correlators. We have not performed this additional  computation.

\section{Discussion}

The mass generation described in this paper presents some similarities with
the Schwinger model in 1+1 dimensions \cite{schwinger}. In that case, even before turning on
any interaction, the peculiar kinematics of 2D space-time implies the presence of a normalizable
massless state composed of a fermion and an anti-fermion. When the gauge coupling is turned on,
the vector field acquires a mass at 1-loop by `eating' that massless bound state.
In our set up it is the kinematics of AdS$_4$ that guarantees the presence of a normalizable
massless scalar state $D(3,0)$. Depending on the boundary conditions,
when the gauge coupling is turned on, the vector boson may eat the bound state and
become massive. The technical difference between the two cases lies in the fact that in
the Schwinger model there is no charge breaking scalar condensate made up of two fermion
fields. This is due to Coleman's theorem \cite{Coleman:1973ci} which establishes
the absence of spontaneous symmetry breaking in 2D field theory. In our $4D$ example, instead, massless
($D(3,0)$) scalar fields have a moduli space of expectation values, determined by boundary conditions.
In this sense the mass generation in AdS$_4$ is qualitatively similar to technicolor theories where
a condensate is responsible for the breaking. However the distinction between the case with and
without condensate is just technical, since in gauge theories the only observable operators are gauge invariant ones.
From a gauge invariant viewpoint the story is the same in the two cases:
starting from a free theory with a massless elementary vector and a massless scalar
`bound' state, a massive vector emerges when the interaction is turned on.

Finally we would like to comment on our results from the standpoint of the AdS/CFT
correspondence \cite{review}. The phenomenon we have studied corresponds to turning on double-trace
(marginal) deformations in the dual CFT$_3$ (see also \cite{porrati} for related
work). A (complex) 4D Weyl fermion $\psi$ quantized to give the $D(E,\frac{1}{2})$
representation corresponds to a (real) 3D fermionic operator $\Psi$ of scaling dimension $E$.
In particular for a massless 4D fermion the dimension of $\Psi$ is $\frac{3}{2}$.
The scalar operator ${\cal O}=\Psi\Psi$ has dimension 3 (in the large N limit) and
represents a double trace marginal deformation. By simple OPE analysis
(like for instance done in ref. \cite{multitrace}) one is indeed convinced that
$\lambda \Psi\Psi$ is exactly marginal. This generalizes to the case of a number
$n$ of fermions $\psi_i$ ($i=1,\dots,n$), corresponding to CFT$_3$ operators $\Psi_i$.
In that case the most general marginal deformation $\Delta {\cal L}_{CFT}=\lambda_{ij} \Psi_i\Psi_j$
is associated to $(n^2+n)/2$ real parameters. This precisely corresponds to the number of real free
parameters in the boundary matrix  $B_{ij}$ on the AdS$_4$ side, although  to derive the mapping between the two sets
some work is needed\cite{porrati}. When the AdS$_4$ gauge group $G\subseteq O(n)$,
the corresponding CFT$_3$ can have global symmetry $G$. The most general marginal  parameters
$\lambda_{ij}$ will in general break $G$ to a subgroup $H$. Corresponding to the vector bosons getting a 1-loop mass in the bulk,
the CFT$_3$ currents in $G/H$ will acquire anomalous dimensions of order $\alpha/4\pi \sim 1/N^2$.
The anomalous dimensions will also depend  on the deformation parameters and vanish continuously at the points
of the $\lambda_{ij}$ space where $G$ invariance is restored. On the other hand,
when $G\not \subseteq O(n)$, corresponding to a chiral gauge theory in AdS, the dual CFT will  be at best invariant
under $G'=G\cap O(n)$. In that case the currents in $G/G'$ will acquire a non-zero $O(1/N^2)$ anomalous
dimension over the entire moduli space.

It is instructive to see in more detail how things work in the simple QED example,
sketching the dual picture of the discussion in subsection 2.1.
The dual CFT contains two fermionic operators that can be packaged into one complex
field $\Psi=\Psi_1+i\Psi_2$, with charge one under the global $U(1)$ symmetry.
The most general double-trace deformation is then
\begin{equation}
\Delta {\cal L}_{CFT}=\lambda_1\Psi^*\Psi+\left (\lambda_2e^{i\theta}\Psi\Psi+{\rm h.c.}\right)
\label{deformation}
\end{equation}
again described by 3 real parameters $(\lambda_1,\,\lambda_2,\,\theta)$.
The phase $\theta$ obviously has no physical consequence, as it can be eliminated
by a $U(1)$ rotation. (This remark applies more generally to the previous discussion:
the physically relevant $\lambda_{ij}$ are determined by modding out by $G$.).
We are left with two physical parameters $\lambda_1$ and $\lambda_2$. When $[\Psi]=\frac{3}{2}$
these parameters are exactly marginal. Of the two, $\lambda_2$ explicitly breaks $U(1)$
and must clearly be consequential: the dual picture of our 4D computation is that the current acquires an anomalous
dimension. The parameter $\lambda_1$ does not break any obvious symmetry of the CFT. What happen here is
clarified by the AdS$_4$ picture. $\lambda_1$ is basically associated to the $\phi_1+\phi_2$ phase in eq.~(\ref{Bgeneral}):
as long as the bulk theory is invariant under the {\it global} chiral symmetry, this phase can be eliminated and bulk
physics remains the same. The CFT interpretation of this phenomenon should be that when $\lambda_1$ is turned on there
exists a field redefinition by which the deformed CFT is shown to be exactly equivalent to the original one.
This peculiarity should correspond to the 3D reflection of 4D global chiral symmetry \cite{porrati}.
Of course one could conceive a 4D theory where, by tuning, the fermions are massless while chirality
is broken by other interactions, for example Yukawa or 4-fermion interactions.
In that situation $\lambda_1$ would parameterize an inequivalent moduli
space of CFTs. Finally, let us consider the case where  a bulk mass is turned on. At the charge preserving points, $\Psi$ has either
dimension $\frac{3}{2}+m$ or $\frac{3}{2}-m$. Consider indeed the second possibility. Eq.~(\ref{deformation}) is
now a relevant deformation. According to the 4D picture, this deformation makes the CFT flow to
 discrete set of inequivalent fixed points. For $\lambda_1\not=0$ and $\lambda_2=0$ electric charge is conserved
and the theory flows to the other possible charge preserving quantization the one where there is a fermion operator $\Psi'$
of dimension $\frac{3}{2}+m$. Instead for $\lambda_1=0$ and $\lambda_2\not=0$ the flow will lead to a new CFT
where the current has a definite $O(1/N^2)$ anomalous dimension.

\vspace{1cm} {\bf Acknowledgments}

We would like to thank Massimo Porrati, Vyacheslav Rychkov, Sergey Sibiryakov, Andrea Wulzer and Alberto Zaffaroni for useful discussions.
Special thanks to Massimo Porrati for prompting us to finish this project. We acknowledge CERN for hospitality during the
final stages of this work. The work of R. R. is supported by the Swiss National Science Foundation
under contract No. 200021-116372.


\end{document}